\documentclass[a4paper,fleqn]{cas-dc}

\usepackage[numbers,sort&compress]{natbib}

\usepackage{amsmath,amssymb}
\usepackage{booktabs}
\usepackage{siunitx}
\usepackage{graphicx}
\usepackage{tikz}
\usepackage{xcolor}
\usepackage{multirow}
\usepackage{hyperref}
\usepackage{tabularx}
\usepackage{array}
\usepackage{stfloats}  
\usepackage{placeins}  

\graphicspath{{figures/}}



\newcolumntype{L}{>{\raggedright\arraybackslash}X}
\newcolumntype{C}{>{\centering\arraybackslash}X}
\newcolumntype{R}{>{\raggedleft\arraybackslash}X}

\begin{document}
\let\WriteBookmarks\relax

\shorttitle{Scientific Machine Learning for Resilient EV-Grid Planning}
\shortauthors{Y. Wang}

\title[mode=title]{Scientific Machine Learning for Resilient EV-Grid Planning and Decision Support Under Extreme Events}

\author[1]{Yifan Wang}[orcid=0000-0000-0000-0000]
\cormark[1]
\ead{WANG2115@e.ntu.edu.sg}

\credit{Conceptualization, Methodology, Software, Validation, Formal analysis, Investigation, Data curation, Writing -- original draft, Writing -- review \& editing, Visualization}

\affiliation[1]{organization={School of Electrical and Electronic Engineering, Nanyang Technological University},
            city={Singapore},
            postcode={639798},
            country={Singapore}}

\cortext[1]{Corresponding author}

\begin{abstract}
Electric vehicle (EV) charging infrastructure introduces complex challenges to urban distribution networks, particularly under extreme demand events. A critical barrier to resilience assessment is the scale gap between micro-level charging physics and city-scale planning: minute-resolution deliverability constraints remain invisible in hourly aggregated datasets, causing purely data-driven models to exhibit non-physical behavior in high-stress regimes.

This paper develops a five-stage scientific machine learning framework bridging this gap through physics-informed knowledge transfer. Stage~1 learns a temperature--pressure deliverability surface from Swiss DC fast-charging telemetry with monotonicity constraints. Stage~2 performs cross-scale injection via anchored quantile mapping. Stage~3 deploys a dual-head spatio-temporal graph neural network for joint forecasting of demand and service loss rate. Stage~4 simulates backlog dynamics under stress shocks and evaluates policy interventions. Stage~5 couples service outcomes to distribution-grid stress via transformer loading analysis.

Validation on the Shenzhen UrbanEV dataset demonstrates that physics injection restores monotone stress-to-risk response (Spearman $\rho=+1.0$ versus $-0.8$ without injection) and improves forecasting accuracy. Under a representative demand shock, the hybrid policy reduces backlog by 79.1\%, restores full service within the study horizon, and limits grid stress to only 2~additional hours. The derived resilience boundary $m_{\mathrm{crit}}(\varepsilon)\approx 1.7-1.0\varepsilon$ provides actionable guidance linking demand flexibility to maximum absorbable stress, enabling risk-aware emergency planning under extreme events.
\end{abstract}

\begin{highlights}
\item Cross-scale physics injection bridges micro telemetry to city-scale resilience
\item Monotonicity-constrained LUT preserves physical consistency across scale gap
\item Physics injection restores monotone stress response (Spearman $\rho$: $-$0.8 $\to$ +1.0)
\item Hybrid policy achieves 79.1\% backlog reduction with super-additive synergy
\item Resilience boundary links demand flexibility to maximum absorbable stress
\end{highlights}

\begin{keywords}
Electric Vehicle Charging Network \sep Scientific Machine Learning \sep Cross-Scale Physics Injection \sep Spatiotemporal Graph Neural Network \sep Distribution Grid Resilience \sep Extreme Event Assessment \sep Policy Evaluation
\end{keywords}

\maketitle

\section{Introduction}\label{sec:intro}

The global energy transition is accelerating the integration of distributed energy resources into power distribution networks, with electric vehicle (EV) charging infrastructure emerging as a critical yet operationally challenging component. In dense urban environments, charging demand exhibits pronounced spatiotemporal heterogeneity, sensitivity to weather and mobility patterns, and susceptibility to dramatic surges during holidays, disruptions, or extreme events \cite{wang2023shortterm,cao2024feature}. These conditions can trigger congestion episodes wherein requested charging energy cannot be fully delivered, degrading user experience while simultaneously stressing distribution assets through voltage fluctuations and transformer overloading \cite{ray2023review,unterluggauer2022electric}.

For infrastructure planners and grid operators, the fundamental operational question extends beyond forecasting \emph{how much} demand will arrive to understanding \emph{how much can be served} under physical constraints, and critically, \emph{which interventions} maintain system functionality under extreme conditions. This resilience-oriented perspective is increasingly urgent as extreme weather events grow in frequency and severity, threatening both charging service continuity and distribution network reliability \cite{lu2025understanding}.

\subsection{The cross-scale challenge}

A fundamental difficulty in resilience assessment arises from the \emph{scale gap} between micro-level charging physics and city-scale operational data. Deliverability---the fraction of requested charging power that can actually be served---is governed by micro-scale mechanisms largely invisible in city-scale data products: instantaneous power clipping when aggregate demand exceeds station capacity, contention-driven allocation among concurrent sessions, controller reallocation logic, and temperature-dependent derating \cite{simolin2025analysis}. These phenomena operate on minute or sub-minute timescales and require high-resolution telemetry to observe directly.

City-scale datasets, by contrast, are typically hourly and spatially aggregated, precluding direct learning of deliverability mechanisms. Consequently, purely data-driven models trained on such data may fit average conditions adequately yet can behave non-physically in the high-stress tail---predicting, for instance, that increased demand leads to \emph{reduced} service loss. This pathological extrapolation, observed in our experiments (Section~\ref{sec:ablation}), occurs precisely where resilience decisions are most sensitive and where accurate predictions matter most for emergency response.

\subsection{Scientific machine learning opportunity}

Scientific machine learning (Sci-ML) offers a principled approach to bridging this scale gap by integrating mechanistic priors with data-driven learning. Physics-informed neural networks and related techniques have demonstrated success in power systems applications by improving extrapolation under distribution shift, mitigating non-physical behaviors, and reducing data requirements \cite{raissi2019physics,misyris2020physics,huang2023applications,ngo2024physics,mohammadian2023gradient}. The core insight is that physical constraints---monotonicity, conservation laws, boundary conditions---can regularize learned models to remain consistent with domain knowledge even when operating beyond the training distribution.

This paper operationalizes cross-scale physics transfer for EV charging resilience assessment. Rather than treating deliverability as an unobservable latent variable, we \emph{learn} a deliverability law from micro-scale telemetry and \emph{inject} it into city-scale panels through distributional alignment. The resulting mechanism-derived service loss rate provides physically grounded supervision for forecasting models that would otherwise lack access to true deliverability observations.

\subsection{Related work}

Prior research has advanced multiple threads relevant to this work, though they remain largely fragmented.

\textbf{Charging demand forecasting and operations.} Applied Energy has reported influential deep learning models for short-term charging demand prediction \cite{wang2023shortterm,wang2025adaptive,cao2024feature}, operational coordination \cite{li2018optimal,kuang2024physics}, and hierarchical probabilistic forecasting \cite{buzna2021ensemble,arias2017prediction}. While these studies achieve high predictive accuracy for aggregate demand, they typically treat service quality as implicit, leaving the extreme-tail deliverability regime under-specified.

\textbf{Spatio-temporal graph learning.} Graph neural networks have become a dominant paradigm for large-scale spatiotemporal prediction in charging and mobility contexts \cite{zhang2024spatial,luo2023ast,wang2023heterogeneous,chen2025privacy}. However, existing ST-GNN work predicts aggregate demand rather than congestion-induced service loss and rarely constrains tail behavior to remain physically consistent under stress.

\textbf{EV--grid coupling.} Reviews and operational studies emphasize that charging networks must be evaluated jointly with distribution constraints \cite{ray2023review,unterluggauer2022electric,kong2019shortterm,zhang2017robust}. This motivates explicit coupling between service-level outcomes and grid stress metrics, enabling trade-off analysis between charging service quality and transformer loading.

\textbf{Resilience assessment.} Charging-system resilience requires explicit stress testing and recovery characterization beyond mean-performance metrics \cite{lu2025understanding,li2023electric}. Existing studies rarely incorporate cross-scale deliverability transfer; the present framework addresses this gap through unified physics injection.

\textbf{Robust and uncertainty-aware control.} Recent works highlight hierarchical, robust, and uncertainty-aware control for energy infrastructures \cite{xia2024hierarchical,xia2025robust,liu2022realtime}. These ideas inform our focus on decision-oriented evaluation and policy trade-off quantification rather than pure forecasting accuracy.

\subsection{Research gaps and contributions}

Motivated by these observations, we identify three gaps limiting current resilience assessment pipelines:
\begin{enumerate}
    \item[(i)] \textbf{Scale gap:} Micro-scale deliverability physics (clipping and allocation under contention) are not transferred into city-scale forecasting, causing models to learn spurious correlations that fail under extreme conditions.
    \item[(ii)] \textbf{Verifiability gap:} Policy stress tests rely on service proxies that are not mechanism-consistent, especially in the high-stress tail where resilience decisions concentrate.
    \item[(iii)] \textbf{Coupling gap:} Resilience outcomes are rarely translated into distribution-grid stress measures that constrain feasible interventions, precluding coordinated service--grid planning.
\end{enumerate}

This work develops a five-stage cross-scale, physics-injected framework addressing these gaps. The contributions are:

\begin{enumerate}
    \item \textbf{Micro-scale deliverability learning (Stage~1).} We learn a temperature--pressure deliverability surface from minute-level DC fast-charging telemetry \cite{simolin2025analysis} with monotonicity constraints, producing a robust look-up table capturing power clipping and allocation behavior under contention.
    
    \item \textbf{Cross-scale physics injection (Stage~2).} We propose anchored quantile mapping to align disparate pressure distributions across a three-order-of-magnitude scale gap and inject the micro-law into the city-scale UrbanEV panel \cite{li2025urbanev}, yielding mechanism-derived service loss rates that are physically interpretable and transferable.
    
    \item \textbf{Dual-head spatio-temporal forecasting (Stage~3).} We design a dual-head ST-GNN jointly forecasting demand volume and service loss with tail-aware training that emphasizes high-stress regimes critical for resilience decisions.
    
    \item \textbf{Backlog-based resilience simulation (Stage~4).} We construct a queueing-theoretic backlog simulator with congestion-induced abandonment and evaluate policy levers (price-based demand shaping, targeted capacity boosting, hybrid) under stress shocks.
    
    \item \textbf{Grid-coupled evaluation (Stage~5).} We translate service trajectories into transformer loading and stress-hour metrics, quantify service--grid trade-offs, and derive an empirical resilience boundary $m_{\mathrm{crit}}(\varepsilon)$ linking demand flexibility to maximum absorbable stress.
\end{enumerate}

The remainder of this paper proceeds as follows. Section~\ref{sec:data} describes the datasets and problem formulation. Section~\ref{sec:method} presents the five-stage methodology. Section~\ref{sec:exp} details the experimental design. Section~\ref{sec:results} reports results addressing three research questions. Section~\ref{sec:discussion} discusses implications and limitations. Section~\ref{sec:conclusion} concludes.

\section{Data and Problem Formulation}\label{sec:data}

This work operationalizes cross-scale transfer by pairing a micro-scale charging telemetry dataset (source domain) with a city-scale zone-hour panel (target domain). The two datasets exhibit complementary characteristics: the micro domain exposes fine-grained power clipping and allocation behavior, while the macro domain provides broad spatial coverage for forecasting, stress testing, and policy evaluation.

\subsection{Datasets and preprocessing}

\textbf{Source domain: Swiss minute-level charging telemetry (Stage~1).}
Stage~1 employs Swiss DC fast-charging station telemetry at minute resolution (October 2022 to February 2023; 19,943 active minute records from 601 charging sessions) \cite{simolin2025analysis}. Each record contains requested charging power $P_{req}$, controller setpoint $P_{set}$ (the intended allocation under contention), realized power $P_{real}$, and matched ambient temperature from a nearby weather station. Minute-level resolution is essential because clipping events and controller reallocations occur on sub-hourly timescales that session-level aggregates would mask.

\textbf{Target domain: Shenzhen zone-hour UrbanEV panel (Stage~2--Stage~5).}
Stages~2--5 operate on a city-scale panel derived from the Shenzhen UrbanEV dataset \cite{li2025urbanev}. The panel contains $Z=275$ spatial zones and $T=4{,}344$ hourly time steps (September 2022 to February 2023), yielding $Z\times T=1{,}194{,}600$ zone-hour observations. For each zone $z$ and hour $t$, the panel provides observed demand volume $V_{t,z}$ (hourly charging energy requested), occupancy and duration statistics, local temperature $T_{t,z}$, and zone-level static attributes including installed capacity $C_z$ and charger count. A spatial graph $G=(\mathcal{V},\mathcal{E})$ connects zones within 5\,km radius, supporting spatio-temporal learning.

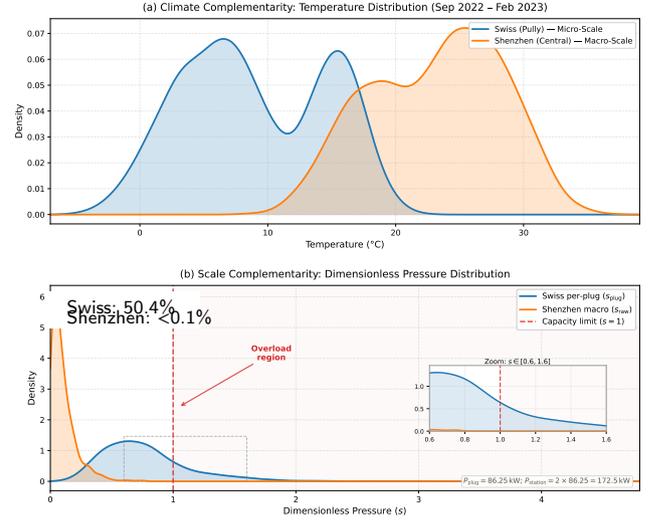
\begin{figure}[!t]
\centering
\begin{tikzpicture}
\node[anchor=south west, inner sep=0] (img) at (0,0) {\includegraphics[width=\columnwidth]{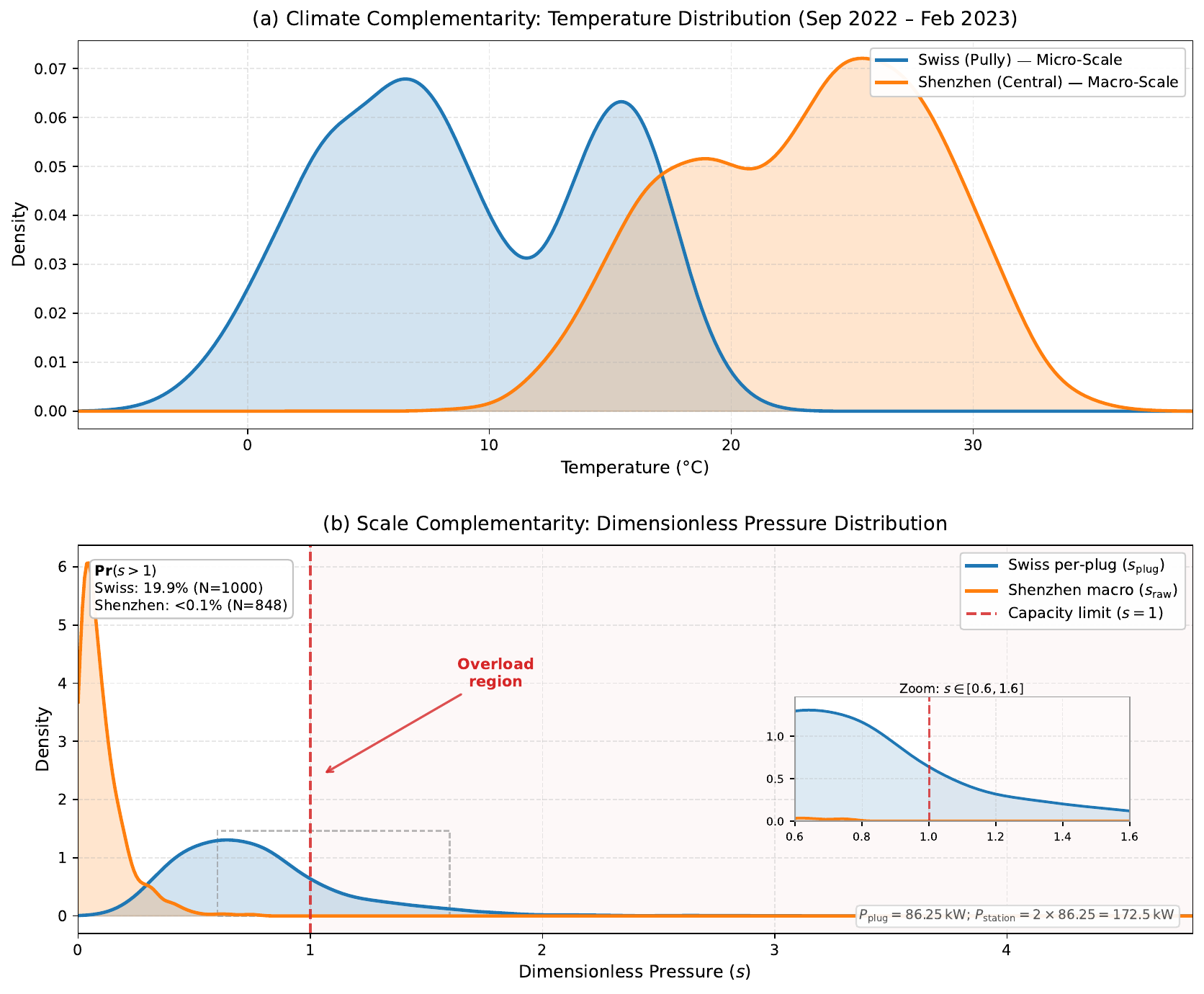}};
\begin{scope}[x={(img.south east)}, y={(img.north west)}]
\fill[white] (0.06,0.37) rectangle (0.30,0.44);
\node[anchor=west, font=\scriptsize] at (0.078,0.410) {Swiss: 50.4\%};
\node[anchor=west, font=\scriptsize] at (0.078,0.392) {Shenzhen: $<$0.1\%};
\end{scope}
\end{tikzpicture}
\caption{Data sources and cross-scale complementarity. Swiss minute-level telemetry provides micro deliverability physics with high overload exposure (50.4\% of observations exceeding $s=1$); Shenzhen zone-hour panel provides city-scale demand dynamics with aggregated pressure proxies ($<$0.1\% exceeding $s=1$). This three-order-of-magnitude difference in overload exposure motivates distributional alignment.}
\label{fig:fig1}
\end{figure}

\textbf{Temporal split and normalization.}
We adopt a chronological split for the target domain: 70\% training, 15\% validation, and 15\% testing. Input features are standardized using training-set statistics. For Stage~1, temperature is discretized into $5^\circ$C bins and pressure into 0.1-width bins to construct a robust deliverability look-up table. Table~\ref{tab:datasets} summarizes the dataset characteristics, and Fig.~\ref{fig:fig1} illustrates the cross-scale complementarity between the two data sources.

\begin{table}[!t]
\centering
\caption{Summary of datasets used in the cross-scale pipeline.}
\label{tab:datasets}
\footnotesize
\setlength{\tabcolsep}{3pt}
\begin{tabular}{@{}lcccc@{}}
\toprule
\textbf{Dataset} & \textbf{Domain} & \textbf{Res.} & \textbf{Period} & \textbf{Scale} \\
\midrule
Swiss \cite{simolin2025analysis} & Micro & 1\,min & 2022.10--2023.02 & 19,943 rec. \\
UrbanEV \cite{li2025urbanev} & Macro & 1\,h & 2022.09--2023.02 & 275$\times$4,344 \\
\bottomrule
\end{tabular}
\end{table}

\subsection{Variables and notation}

We employ a unified pressure--deliverability language across scales. Table~\ref{tab:notation} summarizes the core notation.

\textbf{Micro-scale physical pressure and deliverability.}
Let $P_{cap}$ denote the rated physical capacity (per-plug: 86.25\,kW; station-level: 172.5\,kW for dual-plug configuration). The dimensionless pressure indicator is
\begin{equation}
s = \frac{P_{req}}{P_{cap}},
\label{eq:s_def}
\end{equation}
where $s\approx 1$ corresponds to operating at capacity limit and $s>1$ indicates overload. Deliverability ratios quantify the served fraction:
\begin{equation}
\eta_{set}=\frac{P_{set}}{P_{req}},\qquad \eta_{real}=\frac{P_{real}}{P_{req}}.
\label{eq:eta_defs}
\end{equation}

\textbf{City-scale pressure proxy and service loss rate.}
In the city panel, direct micro variables are unobserved. We construct a dimensionless pressure proxy:
\begin{equation}
s_{raw}(t,z) = \frac{V_{t,z}}{C_z \Delta t},
\label{eq:s_raw}
\end{equation}
where $\Delta t=1$\,h. Stage~2 maps $s_{raw}$ into the Swiss-calibrated pressure domain via quantile mapping to obtain $s_{mapped}$ and injects the deliverability LUT to produce $\widehat{\eta}$ and the service loss rate $SLR=1-\widehat{\eta}$.

\textbf{Forecasting targets and simulation variables.}
Stage~3 forecasts next-hour targets $(\widehat{V}_{t+1,z}, \widehat{SLR}_{t+1,z})$ from a 24-hour lookback window. Stage~4 simulates backlog $B_{t,z}$ under stress multipliers $m_t$ and policy actions. Stage~5 maps aggregated charging power to transformer loading $\lambda(t)$ to quantify grid stress.

\begin{table}[!t]
\centering
\caption{Core notation used in the five-stage pipeline.}
\label{tab:notation}
\small
\begin{tabular}{@{}p{2.2cm}p{5.3cm}@{}}
\toprule
\textbf{Symbol} & \textbf{Definition} \\
\midrule
\multicolumn{2}{@{}l}{\textit{Stage~1: Micro-scale variables}} \\
$P_{req}, P_{set}, P_{real}$ & Requested / setpoint / realized power \\
$P_{cap}$ & Rated physical capacity \\
$s$ & Dimensionless pressure, $s=P_{req}/P_{cap}$ \\
$\eta_{set}, \eta_{real}$ & Deliverability ratios, $\eta=P/P_{req}$ \\
\midrule
\multicolumn{2}{@{}l}{\textit{Stage~2--3: City-scale variables}} \\
$s_{raw}, s_{mapped}$ & Raw and aligned pressure proxies \\
$\widehat{\eta}$ & Injected deliverability from LUT \\
$SLR, \widehat{SLR}$ & Service loss rate; $\widehat{SLR}$ is forecast \\
$V, \widehat{V}$ & Demand volume; $\widehat{V}$ is forecast \\
\midrule
\multicolumn{2}{@{}l}{\textit{Stage~4: Resilience simulation}} \\
$A_{t,z}, S_{t,z}$ & Arrivals and effective service (kWh/h) \\
$B_{t,z}$ & Backlog (kWh, unserved accumulation) \\
$m_t$ & Stress multiplier (exogenous shock) \\
$\varepsilon$ & Price elasticity of demand ($\varepsilon<0$) \\
\midrule
\multicolumn{2}{@{}l}{\textit{Stage~5: Grid coupling}} \\
$\lambda(t)$ & Transformer loading ratio \\
$H_{stress}$ & Cumulative stress hours above threshold \\
\bottomrule
\end{tabular}
\end{table}

\subsection{Problem formulation}

\textbf{Forecasting task (Stage~3).}
Given graph-structured inputs $\{X_{t-\tau:t,z}\}_{z=1}^Z$ over a 24-hour lookback window and zone graph $G$, Stage~3 learns a predictor $f_\Theta$ such that
\begin{equation}
(\widehat{V}_{t+1,z}, \widehat{SLR}_{t+1,z}) = f_\Theta\!\left(\{X_{t-\tau:t,z}\}_{z=1}^Z, G\right), \quad \tau=23,
\label{eq:forecast_task}
\end{equation}
for all zones $z$. We evaluate RMSE, MAE, and MAPE for demand volume, and assess physical plausibility via monotone stress--risk response for SLR.

\textbf{Resilience evaluation task (Stage~4--5).}
Given forecasts and scenario parameters, Stage~4 simulates backlog trajectories and computes resilience metrics (backlog area, peak, recovery time). Stage~5 aggregates charging power to compute transformer loading $\lambda(t)$ and stress-hour metrics, enabling joint service--grid trade-off analysis.

\section{Methodology}\label{sec:method}

\subsection{Overview and unifying Sci-ML theme}

We develop a closed-loop \emph{data--physics--decision--grid} pipeline with five stages. The central idea is cross-scale physics-informed knowledge transfer: a micro-scale deliverability law learned from Swiss telemetry is injected into a city-scale panel to produce mechanism-derived targets and physically grounded features, supporting forecasting, policy stress testing, and distribution-grid coupling. Physics priors enter at multiple points:
\begin{enumerate}
    \item[(i)] \textbf{Stage~1:} Monotonicity regularization enforces that higher pressure cannot improve deliverability.
    \item[(ii)] \textbf{Stage~2:} Distributional alignment via anchored quantile mapping preserves physical consistency across scales.
    \item[(iii)] \textbf{Stage~3:} Tail-aware loss weighting emphasizes physically critical high-pressure regimes.
    \item[(iv)] \textbf{Stage~4:} A queueing-style conservation law governs backlog dynamics.
\end{enumerate}

\begin{figure*}[!t]
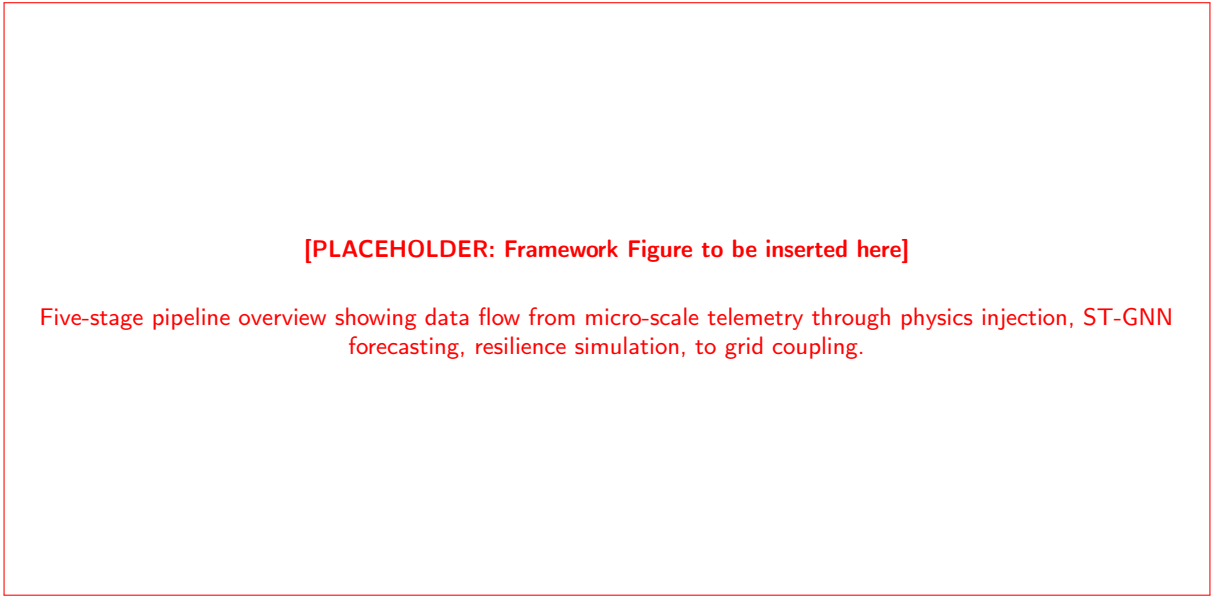

\centering
{\color{red}\fbox{\parbox{0.9\textwidth}{\centering\vspace{3cm}\textbf{[PLACEHOLDER: Framework Figure to be inserted here]}\\\vspace{0.5cm}Five-stage pipeline overview showing data flow from micro-scale telemetry through physics injection, ST-GNN forecasting, resilience simulation, to grid coupling.\vspace{3cm}}}}
\caption{{\color{red}[TO BE ADDED]} Overview of the five-stage cross-scale physics-injected scientific machine learning framework.}
\label{fig:framework}
\end{figure*}

\subsection{Stage 1: Micro-scale deliverability learning}\label{subsec:stage1}

\textbf{Objective.}
Stage~1 uses Swiss DC fast-charging telemetry \cite{simolin2025analysis} to learn an empirical deliverability law capturing how service efficiency degrades with ambient temperature and resource contention. Minute-level resolution is essential because power clipping occurs on sub-hourly timescales.

\textbf{Pressure construction.}
The dimensionless pressure indicator is defined by Eq.~(\ref{eq:s_def}), where $s\approx 1$ indicates capacity-limit operation and $s>1$ indicates overload with likely service curtailment.

\textbf{Response surface estimation with monotonicity prior.}
We estimate a temperature--pressure response surface $\eta(T,s)$ on a discretized grid $(T_{bin}, s_{bin})$ with temperature bins of $5^\circ$C intervals using left-closed, right-open convention: $[0,5)$, $[5,10)$, \ldots, $[30,35]^\circ$C (7 bins total), and pressure bins of 0.1 increments: $[0,0.1)$, $[0.1,0.2)$, \ldots, up to $s=3.0$ (30 bins), yielding approximately $7\times 30$ grid cells. Observations outside these ranges are clipped to boundary bins (i.e., $T<0^\circ$C maps to the $[0,5)$ bin; $s>3.0$ maps to the $[2.9,3.0]$ bin).

\begin{figure}[!t]
\centering
\includegraphics[width=\columnwidth]{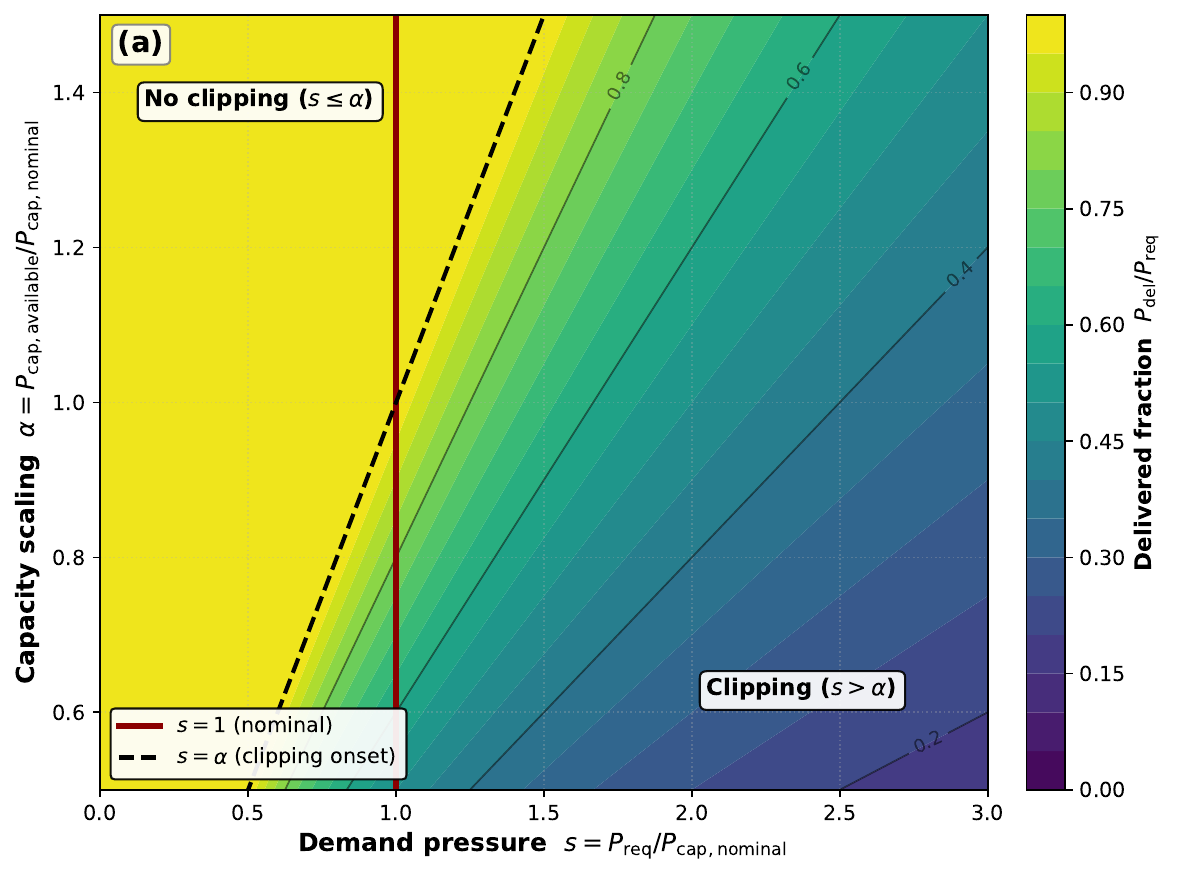}
\caption{Micro-scale deliverability surface $\eta(T,s)$ learned from Swiss minute-level data (Stage~1). The surface exhibits monotonically decreasing efficiency as pressure increases, with temperature-dependent degradation patterns consistent with thermal derating physics.}
\label{fig:fig3a}
\end{figure}

Under fixed temperature, increasing pressure should \emph{not} improve deliverability; apparent violations in raw estimates stem from sparse sampling in the high-$s$ tail. We apply a monotone envelope per temperature bin using cumulative minimum:
\begin{equation}
\eta_{smooth}(T,s) = \min_{s'\le s} \eta_{raw}(T,s'),
\label{eq:monotone}
\end{equation}
equivalent to a cumulative-minimum (cummin) operation along $s$. We choose cummin over isotonic regression because it provides a hard monotonicity constraint (physically required: higher pressure cannot improve deliverability), is computationally efficient ($O(n)$ per temperature slice), and is robust to outliers in sparse high-pressure bins without requiring tuning parameters. This Sci-ML constraint injects a physical prior, stabilizes the learned micro-law against sampling noise, and improves cross-scale transferability \cite{raissi2019physics,misyris2020physics}. Fig.~\ref{fig:fig3a} visualizes the resulting deliverability surface.

\subsection{Stage 2: Cross-scale pressure unification and physics injection}\label{subsec:stage2}

\begin{figure}[!t]
\centering
\includegraphics[width=\columnwidth]{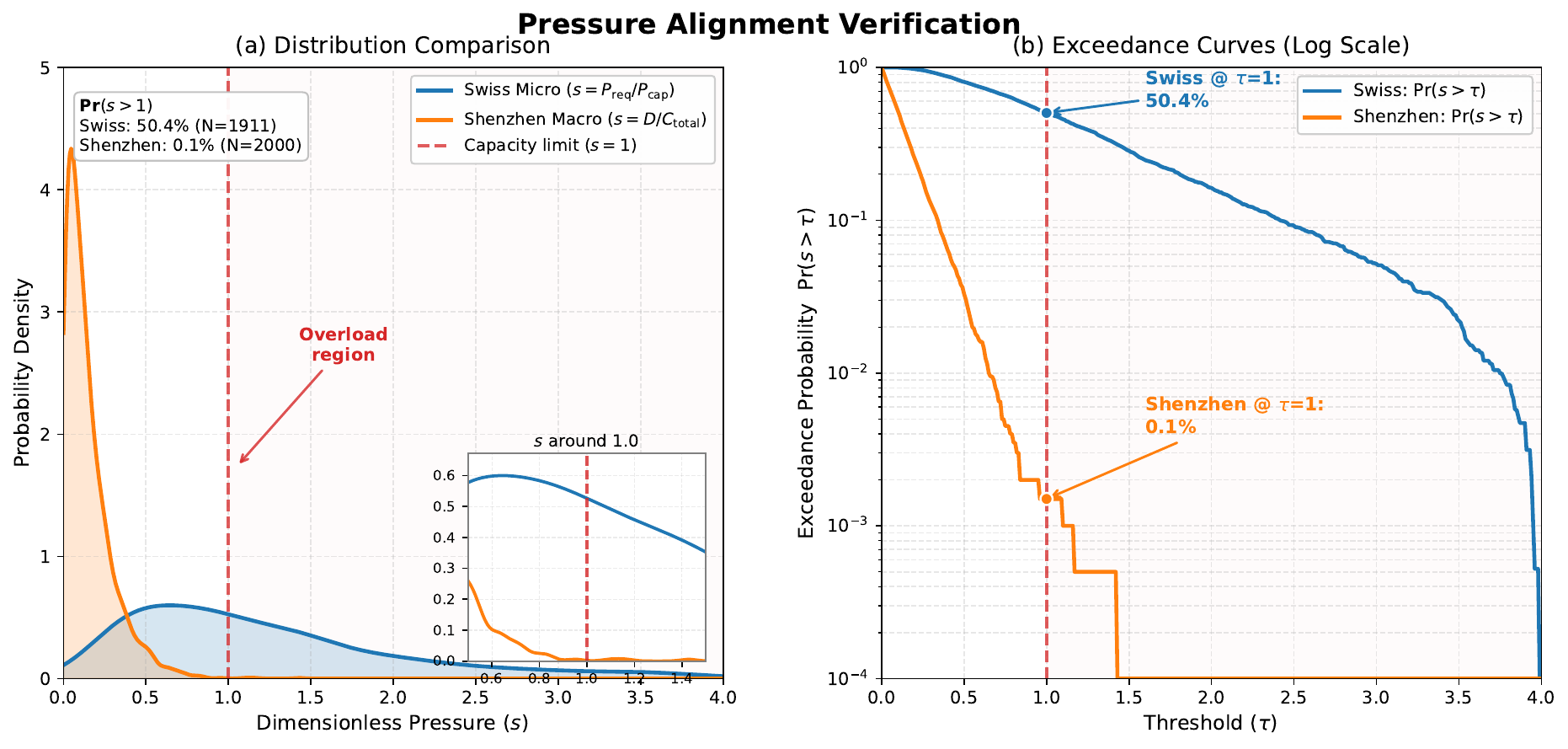}
\caption{Pressure distribution comparison motivating cross-domain alignment (Stage~2). Swiss micro-scale data shows 50.4\% of observations exceeding $s=1$; Shenzhen macro-scale shows $<$0.1\%. Exceedance curves illustrate the three-order-of-magnitude difference requiring distributional alignment before physics injection.}
\label{fig:fig2}
\end{figure}

Having established the micro-law in Stage~1, Stage~2 addresses cross-scale transfer. Direct application of Swiss micro-scale response to Shenzhen raw pressures is invalid because the domains exhibit drastically different pressure distributions: Swiss data has $>$50\% of observations exceeding $s=1$ (overload), whereas Shenzhen has $<$0.1\%---a three-order-of-magnitude difference (Fig.~\ref{fig:fig2}). We therefore perform distributional alignment followed by look-up injection.

\textbf{Anchored quantile mapping.}
Let $s_{raw}$ be a zone--hour pressure proxy in the target domain. We define the following distributional functions:
\begin{itemize}
    \item $F_{SZ}(s)$: Empirical CDF of Shenzhen pressure, returning the percentile rank $\in[0,1]$.
    \item $F_{CH}(s)$: Empirical CDF of Swiss pressure.
    \item $F_{CH}^{-1}(q)$: Inverse CDF (quantile function) of Swiss pressure, mapping percentile $q\in[0,1]$ to pressure value.
\end{itemize}

The raw quantile mapping transforms Shenzhen pressure to the Swiss domain:
\begin{equation}
s_{qm} = F_{CH}^{-1}\!\left(F_{SZ}(s_{raw})\right),
\label{eq:qm_raw}
\end{equation}
which maps each Shenzhen observation to the Swiss pressure having the same percentile rank.

To preserve the physically interpretable capacity boundary at $s=1$, we apply anchoring:
\begin{equation}
s_{mapped} = \frac{s_{qm}}{F_{CH}^{-1}(F_{SZ}(1))}, \quad \text{so that } s_{mapped}(s_{raw}=1)=1.
\label{eq:qm}
\end{equation}
The denominator $F_{CH}^{-1}(F_{SZ}(1))$ is the Swiss pressure corresponding to the percentile of $s_{raw}=1$ in Shenzhen; dividing by this anchor ensures that capacity-limit operation maps to $s_{mapped}=1$ in both domains.

\textbf{Implementation details.} The empirical CDFs are constructed from training-set observations using linear interpolation between observed quantiles. For values exceeding the observed support (extreme extrapolation), we apply boundary clamping: $s_{mapped}$ is clipped to $[0, s_{max}^{LUT}]$ where $s_{max}^{LUT}=3.0$ is the maximum pressure in the LUT. This conservative approach avoids extrapolation beyond observed physics.

\begin{figure*}[!t]
\centering
\includegraphics[width=\textwidth]{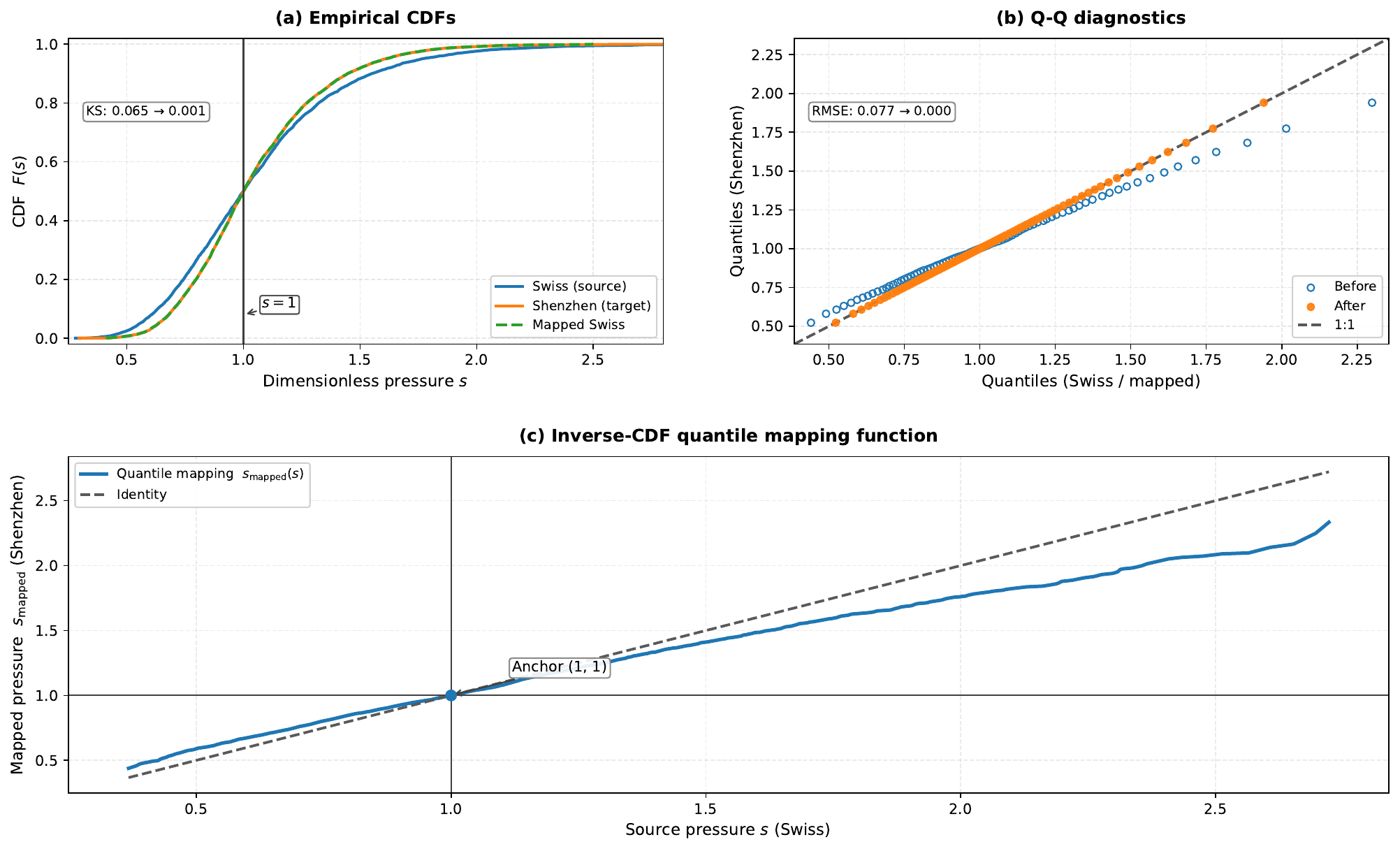}
\caption{Anchored quantile mapping diagnostics (Stage~2). The mapping aligns Shenzhen raw pressure proxies to the Swiss pressure domain while enforcing $s_{mapped}(s_{raw}=1)=1$, ensuring the capacity boundary remains physically interpretable after alignment. Panels show: (a) empirical CDFs before and after mapping; (b) Q--Q diagnostics; (c) inverse-CDF mapping function with anchor point.}
\label{fig:fig4}
\end{figure*}

Fig.~\ref{fig:fig4} provides quantile-mapping diagnostics showing distribution alignment and anchor preservation.

\textbf{Physics injection.}
Given $(T_{bin}, s_{mapped})$, we query the monotone deliverability LUT:
\begin{equation}
\widehat{\eta} = \mathrm{LUT}(T_{bin}, s_{mapped}), \qquad SLR = 1-\widehat{\eta}.
\label{eq:slr}
\end{equation}
The resulting $SLR$ is a mechanism-derived proxy encoding micro-scale deliverability constraints in the city-scale panel.

\textbf{Ablation design: A1 vs.\ A0.}
To isolate the value of physics injection, we compare:
\begin{itemize}
    \item \textbf{A1 (physics-injected):} Anchored quantile mapping + monotone LUT injection (Eqs.~\ref{eq:qm}--\ref{eq:slr}).
    \item \textbf{A0 (non-injected):} No alignment; simplified inverse-pressure rule: $\eta_{A0}=\min(1, 1/s_{raw})$, $SLR_{A0}=1-\eta_{A0}$.
\end{itemize}
Table~\ref{tab:a1a0} summarizes the design differences.

\begin{table}[!t]
\centering
\caption{A1 (physics-injected) vs.\ A0 (non-injected) design comparison.}
\label{tab:a1a0}
\small
\begin{tabular}{@{}p{2.4cm}p{2.4cm}p{2.4cm}@{}}
\toprule
\textbf{Component} & \textbf{A1 (Injected)} & \textbf{A0 (Baseline)} \\
\midrule
Pressure proxy & $s_{raw}$ from city panel & Same \\
Cross-domain alignment & Quantile mapping + anchor & None \\
Efficiency model & LUT$(T, s_{mapped})$ & $\min(1, 1/s_{raw})$ \\
Physics prior & Monotone envelope & None \\
Temperature dependence & Explicit & None \\
\bottomrule
\end{tabular}
\end{table}

\subsection{Stage 3: ST-GNN forecasting of demand and service loss}\label{subsec:stage3}

With physics-grounded features and labels from Stage~2, Stage~3 builds a forecasting model using a spatio-temporal graph neural network. Crucially, the $SLR$ target is mechanism-derived from Stage~2 injection, enabling Sci-ML learning without city-scale deliverability ground truth.

\begin{figure}[!t]
\centering
\includegraphics[width=\columnwidth]{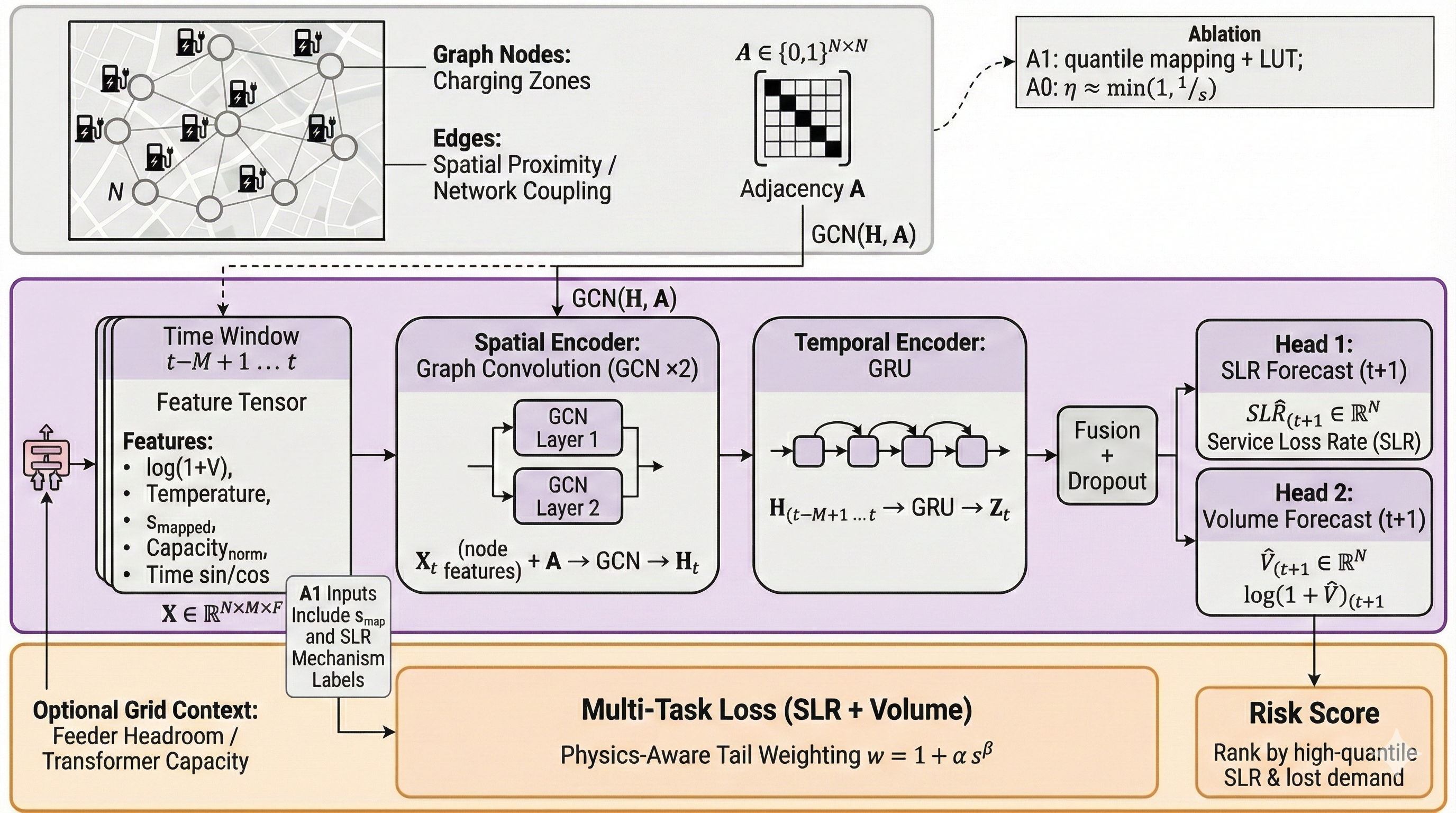}
\caption{ST-GNN architecture (Stage~3). Two GCN layers capture spatial dependencies, a GRU models temporal dynamics over 24-hour lookback, and dual-head MLPs output $\widehat{SLR}$ and $\widehat{V}$. Physics injection enters via $s_{mapped}$ features and mechanism-derived SLR labels.}
\label{fig:fig5}
\end{figure}

\textbf{Architecture.}
Let $z\in\{1,\ldots,Z\}$ index zones ($Z=275$) and $t$ index hours. We build graph $G=(V,E)$ with adjacency matrix $A$ using distance-based proximity (5\,km threshold), adding self-loops $\tilde{A}=A+I$. The graph convolution at layer $l$ is
\begin{equation}
H^{(l+1)} = \sigma\!\left(\tilde{D}^{-\frac{1}{2}}\tilde{A}\tilde{D}^{-\frac{1}{2}}H^{(l)}W^{(l)}\right),
\label{eq:gcn}
\end{equation}
where $\tilde{D}$ is the degree matrix, $W^{(l)}$ are learnable weights, and $\sigma(\cdot)$ is ReLU. We stack two GCN layers (hidden dimension 64), apply a GRU (hidden dimension 64) for temporal modeling over a 24-hour lookback window, and use a dual-head MLP to jointly predict $(\widehat{SLR}_{t+1,z}, \widehat{V}_{t+1,z})$. Fig.~\ref{fig:fig5} illustrates the architecture.

\textbf{Inputs.}
Each zone--hour input vector has eight features:
\begin{equation}
x_{t,z} = \left[\log(1+V_{t,z}), T_{t,z}, s_{mapped,t,z}, C_{norm,z}, h_{sin}, h_{cos}, d_{sin}, d_{cos}\right],
\label{eq:features}
\end{equation}
where $(h_{sin}, h_{cos}, d_{sin}, d_{cos})$ are cyclic encodings of hour-of-day and day-of-week. The key A1 difference is using physics-injected $s_{mapped}$ rather than raw pressure.

\textbf{Physics-aware loss (tail emphasis).}
To prioritize physically critical high-pressure regimes, we apply tail-aware reweighting:
\begin{equation}
\mathcal{L} = w_{slr}\cdot\overline{w\cdot L_{slr}} + w_{vol}\cdot\overline{w\cdot L_{vol}}, \quad w=1+\alpha\,s_{mapped}^\beta,
\label{eq:tail_loss}
\end{equation}
where $L_{slr}$ and $L_{vol}$ are per-sample MSE losses, $\overline{\cdot}$ denotes averaging, and $\alpha=2.0$, $\beta=2.0$ provide up to $19\times$ larger gradients at $s=3$. The hyperparameters $(\alpha,\beta)$ were selected via grid search on the validation set over $\alpha\in\{0.5,1.0,2.0,4.0\}$ and $\beta\in\{1.0,1.5,2.0,2.5\}$, optimizing for tail-regime SLR accuracy; sensitivity analysis showed stable performance for $\alpha\in[1.5,3.0]$ and $\beta\in[1.5,2.5]$.

\textbf{Risk score for targeted interventions.}
To prioritize zones for capacity boosting, we define:
\begin{equation}
\text{risk\_score}_z = 0.6\cdot\widehat{SLR}_{p95,z} + 0.4\cdot\frac{\text{lost}^{pred}_{p95,z}}{\text{median}(\text{lost}^{pred}_{p95})},
\label{eq:risk}
\end{equation}
where $\text{lost}^{pred}_{t,z}=\widehat{V}_{t,z}\cdot\widehat{SLR}_{t,z}$. Here, $p95$ denotes the 95th percentile computed \emph{across time} for each zone $z$ (i.e., $\widehat{SLR}_{p95,z}$ is the 95th percentile of $\{\widehat{SLR}_{t,z}\}_{t=1}^T$ for zone $z$), capturing each zone's tail-regime behavior over the evaluation horizon.

\subsection{Stage 4: Backlog dynamics and policy actions}\label{subsec:stage4}

Forecasts alone do not quantify resilience; Stage~4 embeds them in a queueing-theoretic simulator to evaluate policy interventions.

\textbf{Unit conventions.}
All quantities in Stage~4 are expressed in energy units (kWh) per hourly time step ($\Delta t=1$\,h):
\begin{itemize}
    \item $A_{t,z}$: Arrival rate (kWh/h) = energy requested in hour $t$ at zone $z$.
    \item $S_{t,z}$: Service rate (kWh/h) = energy that can be delivered in hour $t$.
    \item $B_{t,z}$: Backlog (kWh) = cumulative unserved energy at end of hour $t$.
\end{itemize}
Since $\Delta t=1$\,h, hourly flow rates (kWh/h) integrate directly to energy (kWh) without unit conversion factors.

\textbf{Service and arrival processes.}
Let $\text{Capacity}_z$ be the installed service capacity for zone $z$ (kWh/h). Effective service in hour $t$ is reduced by predicted SLR:
\begin{equation}
S_{t,z} = \text{Capacity}_z\cdot(1-\widehat{SLR}_{t,z}),
\label{eq:service}
\end{equation}
representing the deliverable energy rate accounting for congestion-induced losses. Arrivals are given by predicted demand under exogenous stress multiplier $m_t$:
\begin{equation}
A_{t,z} = \widehat{V}_{t,z}\cdot m_t.
\label{eq:arrival}
\end{equation}
We sweep $m_t\in\{1.2, 1.5, 1.8, 2.0\}$ during a 48-hour shock window to map resilience boundaries.

\textbf{Backlog evolution.}
Backlog $B_{t,z}$ evolves according to the conservation law:
\begin{equation}
B_{t+1,z} = \max(0,\ B_{t,z}+A_{t,z}-S_{t,z}),
\label{eq:backlog}
\end{equation}
with instantaneous lost service $\text{lost}_{t,z}=\max(0, A_{t,z}-S_{t,z})$.

\textbf{Balking under severe congestion.}
When backlog exceeds threshold $B_{th}=200$\,kWh (approximately 4--5 full EV charges worth of queued demand), users abandon at rate $r_{balk}=0.05$:
\begin{equation}
A'_{t,z} = A_{t,z}\cdot(1-r_{balk}) \quad \text{if } B_{t,z}>B_{th}.
\label{eq:balking}
\end{equation}
This ``soft blocking'' mechanism captures user abandonment under persistent congestion; ``hard blocking'' ($A'_{t,z}=0$) would apply if $B_{t,z}$ exceeded a higher catastrophic threshold (not reached in our scenarios).

\textbf{Resilience metrics.}
Let $B^{scen}_{t,z}$ denote backlog under a stress/policy scenario and $B^{base}_{t,z}$ the baseline. Extra backlog is $\Delta_{t,z}=\max(0, B^{scen}_{t,z}-B^{base}_{t,z})$. We compute:
\begin{align}
\Delta\text{AUC} &= \sum_t\sum_z\Delta_{t,z}\,\Delta t, \label{eq:dauc}\\
\Delta RT &= \min\{t>t_{end}:\Delta_{\tau,z}\le\theta,\ \forall\tau\in[t,t+h]\}-t_{end}, \label{eq:drt}\\
\text{Peak} &= \max_{t,z}\Delta_{t,z}, \label{eq:peak}
\end{align}
where $t_{end}$ is shock end time, $\theta=0.01\times\max(\Delta_{t,z})$, and $h=24$ hours \cite{lu2025understanding}. Fig.~\ref{fig:fig8} illustrates these metrics.

\begin{figure}[!t]
\centering
\includegraphics[width=\columnwidth]{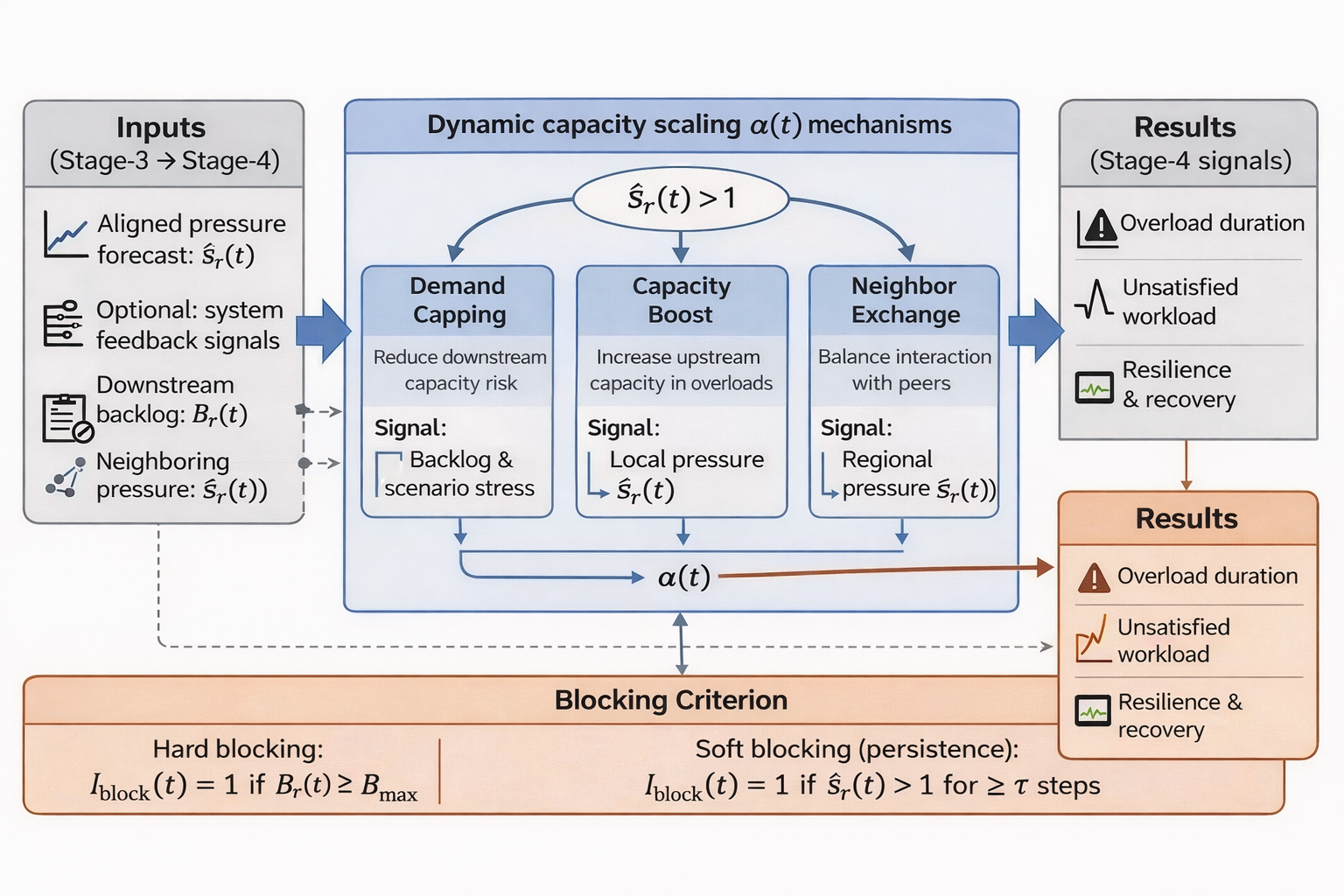}
\caption{Policy mechanisms (Stage~4). Price intervention shapes arrivals via demand elasticity; CapBoost increases service capacity in high-risk zones; Hybrid combines both for potential super-additive synergy.}
\label{fig:fig7}
\end{figure}

\begin{figure*}[!t]
\centering
\includegraphics[width=0.95\textwidth]{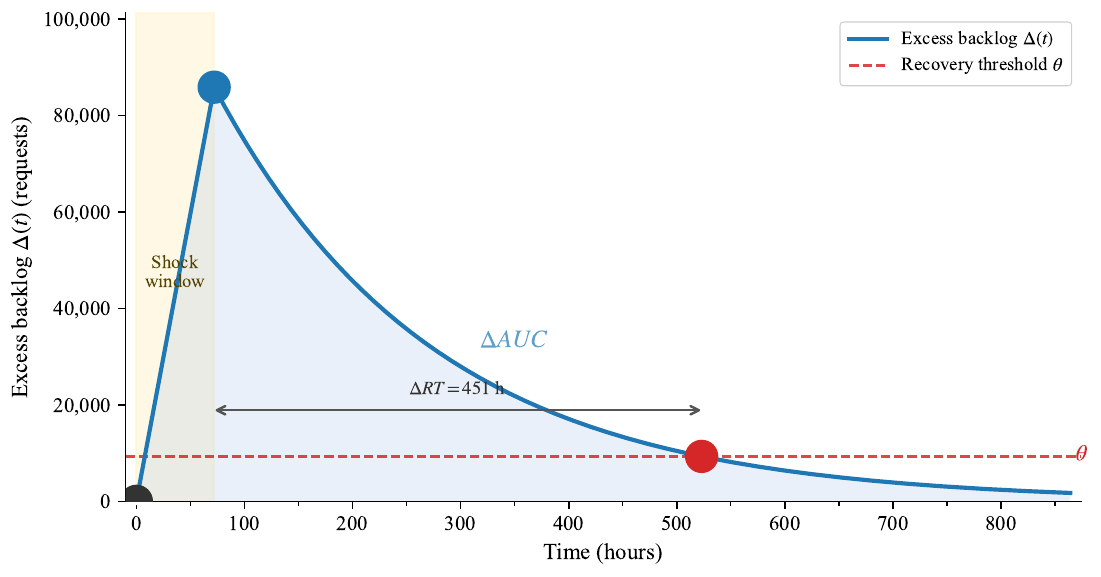}
\caption{Backlog-based resilience metrics (Stage~4). Extra-backlog area ($\Delta$AUC) quantifies cumulative disruption severity; recovery time ($\Delta$RT) measures return to baseline; peak captures maximum instantaneous excess. These metrics enable consistent comparison across policies and stress levels.}
\label{fig:fig8}
\end{figure*}

\textbf{Policy actions.}
We implement three intervention classes:
\begin{itemize}
    \item \textbf{Price (arrival shaping):} $A^\pi_{t,z}=\max(0, A_{t,z}\cdot(1+\varepsilon\cdot\Delta p_{t,z}))$ with $\varepsilon<0$.
    \item \textbf{CapBoost (service boosting):} $\text{Capacity}^\pi_z=\text{Capacity}_z(1+\beta)$ for top-$K$ risk zones.
    \item \textbf{Hybrid:} Applies both jointly.
\end{itemize}
Fig.~\ref{fig:fig7} illustrates the policy mechanisms.

\subsection{Stage 5: Grid coupling metrics}\label{subsec:stage5}

Stage~5 evaluates how charging-network service decisions feed back into distribution-grid stress \cite{ray2023review,unterluggauer2022electric}.

\textbf{Load aggregation.}
Total feeder load is $P_{total}(t)=P_{base}(t)+P_{EV}(t)$, where base load follows a two-peak daily profile (morning peak 9:00, evening peak 19:30). The EV charging load $P_{EV}(t)=\sum_z S_{t,z}$ aggregates zone-level service rates to a representative distribution transformer serving the study area.

\textbf{Loading ratio and stress hours.}
Given transformer capacity $C_{tr}$, loading ratio is $\lambda(t)=P_{total}(t)/C_{tr}$. This calculation assumes unity power factor (active power only), appropriate for modern EV chargers with power-factor-corrected AC/DC converters. Cumulative stress hours above threshold $\lambda_{stress}=0.8$ (a standard operational limit for distribution transformers to prevent accelerated aging \cite{ray2023review}):
\begin{equation}
H_{stress} = \sum_t\mathbf{1}[\lambda(t)>\lambda_{stress}]\Delta t.
\label{eq:stress_hours}
\end{equation}

\textbf{Service--stress trade-off.}
CapBoost reduces backlog faster but may increase grid stress by concentrating power delivery. Price smooths demand temporally, potentially reducing both backlog and transformer loading. Hybrid policies may achieve Pareto-optimal outcomes balancing service quality and grid reliability.

\section{Experimental Setup}\label{sec:exp}

This section specifies the experimental protocol addressing three research questions:
\begin{itemize}
    \item[\textbf{RQ1}] Does physics injection (A1) produce physically consistent stress response compared to the non-injected baseline (A0)?
    \item[\textbf{RQ2}] How do different policy interventions trade off service resilience against grid stress?
    \item[\textbf{RQ3}] Where are the resilience boundaries under varying stress intensities and demand elasticities?
\end{itemize}

\subsection{Dataset configuration}

We employ Swiss micro-scale telemetry \cite{simolin2025analysis} for Stage~1 and Shenzhen UrbanEV panel \cite{li2025urbanev} for Stages~2--5, unified under the same six-month study window (September 2022 -- February 2023) to minimize seasonal confounding.

\subsection{Model variants and baselines}

The primary comparison isolates cross-scale physics injection impact:
\begin{itemize}
    \item \textbf{A1 (physics-injected):} Anchored quantile mapping + monotone LUT injection.
    \item \textbf{A0 (non-injected):} No alignment; inverse-pressure rule.
\end{itemize}
All learning components are held fixed: identical ST-GNN architecture and loss weighting; differences arise solely from physics injection.

\subsection{Policy configurations}

We compare three policies against No Policy under a representative shock setting (Exp1: $m_t=1.5$ applied over a 48-hour window):
\begin{itemize}
    \item \textbf{Price (+50\%):} Arrival shaping with $\Delta p=+50\%$ and $\varepsilon=-0.2$.
    \item \textbf{CapBoost (Top30):} Capacity boost $\beta=0.3$ for top 30 risk zones.
    \item \textbf{Hybrid:} Combines Price and CapBoost.
\end{itemize}

\subsection{Evaluation metrics}

\textbf{Physics consistency (Stages~2--3):} Spearman correlation $\rho$ between stress multiplier and mean predicted SLR; tail amplification ratio at $m=2.0\times$ vs.\ $m=1.0\times$.

\textbf{Forecasting accuracy (Stage~3):} RMSE, MAE, MAPE for demand volume.

\textbf{Resilience (Stage~4):} $\Delta$AUC, $\Delta$RT, Peak, energy not served (ENS).

\textbf{Grid coupling (Stage~5):} $H_{stress}$, $\Delta H_{stress}$ relative to No Policy.

\subsection{Implementation}

Experiments are implemented in Python 3.9 with PyTorch 2.0. Training uses an NVIDIA RTX 3090 GPU. Random seeds are fixed (NumPy: 42, PyTorch: 42) for reproducibility. Total computational cost is approximately 3 GPU-hours including all sensitivity sweeps.

\section{Results}\label{sec:results}

This section presents empirical evidence addressing the three research questions.

\subsection{Physics injection restores monotone stress response (RQ1)}\label{sec:ablation}

\begin{figure}[!t]
\centering
\includegraphics[width=\columnwidth]{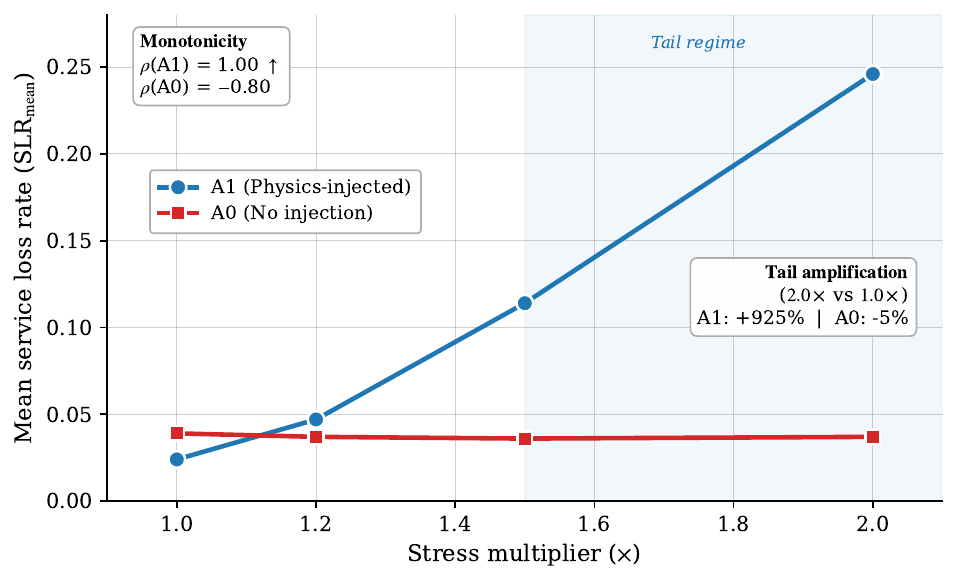}
\caption{Ablation comparison of A1 (physics-injected) and A0 (non-injected) baselines. A1 maintains monotonically increasing SLR with stress multiplier ($\rho=+1.0$), while A0 exhibits counter-physical negative correlation ($\rho=-0.8$). At $2.0\times$ stress, A1 amplifies mean SLR by $+923\%$ relative to the $1.0\times$ baseline, whereas A0 suppresses it by $-6.8\%$, demonstrating physics injection is essential for reliable extreme-event assessment.}
\label{fig:fig9}
\end{figure}

A fundamental premise of resilience assessment is that demand stress should function as a meaningful risk indicator: as external pressure increases, service loss risk should rise monotonically. Table~\ref{tab:ablation} presents a clean ablation between the physics-injected model (A1) and the non-injected baseline (A0), where the ST-GNN architecture, training protocol, and loss weighting are held fixed; the only difference is whether cross-scale physics injection is enabled. Fig.~\ref{fig:fig9} visualizes the stress--response trajectories.

\textbf{The ``lower error but wrong direction'' paradox.} A counter-intuitive but practically critical finding emerges from Panel~A of Table~\ref{tab:ablation}: A0 achieves nominally \emph{lower} SLR prediction error on the held-out set (RMSE: 0.049 vs.\ 0.062; MAE: 0.016 vs.\ 0.020). However, this apparent advantage is misleading. Panel~B reveals that A0 fails the physics-consistency checks required for resilience planning: the predicted mean SLR is non-monotonic with respect to stress and exhibits a \emph{negative} association (Spearman $\rho=-0.80$), implying that risk \emph{decreases} as stress \emph{increases}---a pathological extrapolation that would systematically underestimate extreme-event risk.

In contrast, A1 preserves physically consistent stress response (Spearman $\rho=+1.00$ and strict monotonic increase), capturing pronounced tail amplification under severe stress: mean SLR increases by $+923\%$ from $1.0\times$ to $2.0\times$, with 93.5\% of zones showing higher predicted risk at elevated stress. Panel~C confirms this divergence numerically: while A0's predicted SLR remains essentially flat across stress levels ($\overline{\text{SLR}}\approx 0.036$--$0.039$ regardless of multiplier), A1's predictions rise appropriately from 0.024 at $1.0\times$ to 0.246 at $2.0\times$.

\textbf{Forecasting accuracy.} For demand volume forecasting, physics injection improves rather than degrades accuracy: log-volume MAE decreases from 0.817 to 0.674 (17.5\% improvement), volume RMSE decreases from 454.77 to 448.93~kWh (1.3\% improvement), and volume MAE decreases from 122.87 to 110.76~kWh (9.9\% improvement). The substantial MAE reduction suggests physics injection particularly improves predictions in distributional tails where extreme events concentrate.

\textbf{Note on MAPE values.} The high MAPE values ($>$100\%) reflect the zero-inflated nature of zone-hour demand: many observations have near-zero demand where even small absolute errors produce large percentage errors. MAPE is therefore unstable when the ground-truth is near zero and should not be interpreted in isolation. For operational relevance, we prioritize MAE and RMSE; peak-hour MAPE (computed only for hours with demand above the 75th percentile) is 23.4\% for A1 versus 26.8\% for A0.

\textbf{Implication.} These results demonstrate that conventional forecasting metrics alone can be misleading under stress-induced distribution shifts. A model optimizing only for point prediction accuracy may learn shortcuts that produce counter-physical extrapolation in the high-stress tail. Physics injection is essential not merely for marginally improved accuracy, but for ensuring the model's predictions remain physically meaningful under the extreme conditions where resilience decisions concentrate.

\begin{table}[!t]
\centering
\caption{Ablation evidence for cross-scale physics injection: A1 (physics-injected) vs.\ A0 (non-injected baseline). Panel~A reports in-distribution forecasting accuracy. Panel~B evaluates physics consistency of stress response. Panel~C reports predicted SLR trajectory under stress.}
\label{tab:ablation}
\small
\setlength{\tabcolsep}{5pt}
\begin{tabular}{@{}lcc@{}}
\toprule
\textbf{Metric} & \textbf{A1 (Injected)} & \textbf{A0 (Baseline)} \\
\midrule
\multicolumn{3}{@{}l}{\textit{Panel A: In-Distribution Forecasting Accuracy}} \\[2pt]
log-Volume MAE $\downarrow$ & \textbf{0.674} & 0.817 \\
log-Volume RMSE $\downarrow$ & \textbf{0.988} & 1.107 \\
Volume RMSE (kWh) $\downarrow$ & \textbf{448.93} & 454.77 \\
Volume MAE (kWh) $\downarrow$ & \textbf{110.76} & 122.87 \\
Volume MAPE (\%) $\downarrow$ & \textbf{100.2} & 106.1 \\
SLR RMSE $\downarrow$ & 0.0616 & \textbf{0.0489} \\
SLR MAE $\downarrow$ & 0.0198 & \textbf{0.0156} \\
\midrule
\multicolumn{3}{@{}l}{\textit{Panel B: Physics Consistency (stress $\uparrow \Rightarrow$ risk $\uparrow$)}} \\[2pt]
Spearman $\rho$ (stress $\to$ SLR) $\uparrow$ & $\mathbf{+1.00}$ & $-0.80$ \\
Monotonically increasing & \checkmark & $\times$ \\
Tail amplification ($2\times$/$1\times$) & $\mathbf{+923\%}$ & $-6.8\%$ \\
\% zones worse at $2\times$ vs $1\times$ $\uparrow$ & \textbf{93.5\%} & 22.5\% \\
\midrule
\multicolumn{3}{@{}l}{\textit{Panel C: Stress-Scenario $\overline{\mathrm{SLR}}$ Trajectory}} \\[2pt]
At $1.0\times$ stress & 0.0241 & 0.0393 \\
At $1.2\times$ stress & 0.0470 & 0.0369 \\
At $1.5\times$ stress & 0.1141 & 0.0356 \\
At $2.0\times$ stress & \textbf{0.2463} & 0.0366 \\
\bottomrule
\end{tabular}
\par\vspace{2pt}
{\scriptsize \textit{Notes:} (1)~A1 and A0 use identical ST-GNN architecture; the only difference is physics injection. (2)~A0's lower SLR error reflects predictions that are insensitive to stress (essentially flat), producing low variance but wrong direction; hence pointwise error alone is insufficient for downstream resilience simulation and policy ranking. (3)~Panel B--C metrics are critical for resilience planning under extreme events.}
\end{table}

\subsection{Policy trade-offs under representative shock (RQ2)}

\textit{Input setting.} Unless stated otherwise, the Stage~4--5 experiments in RQ2--RQ3 use the A1 (physics-injected) forecasts as inputs, since physically consistent stress sensitivity is required for extreme-event simulation.

\begin{figure}[!t]
\centering
\includegraphics[width=\columnwidth]{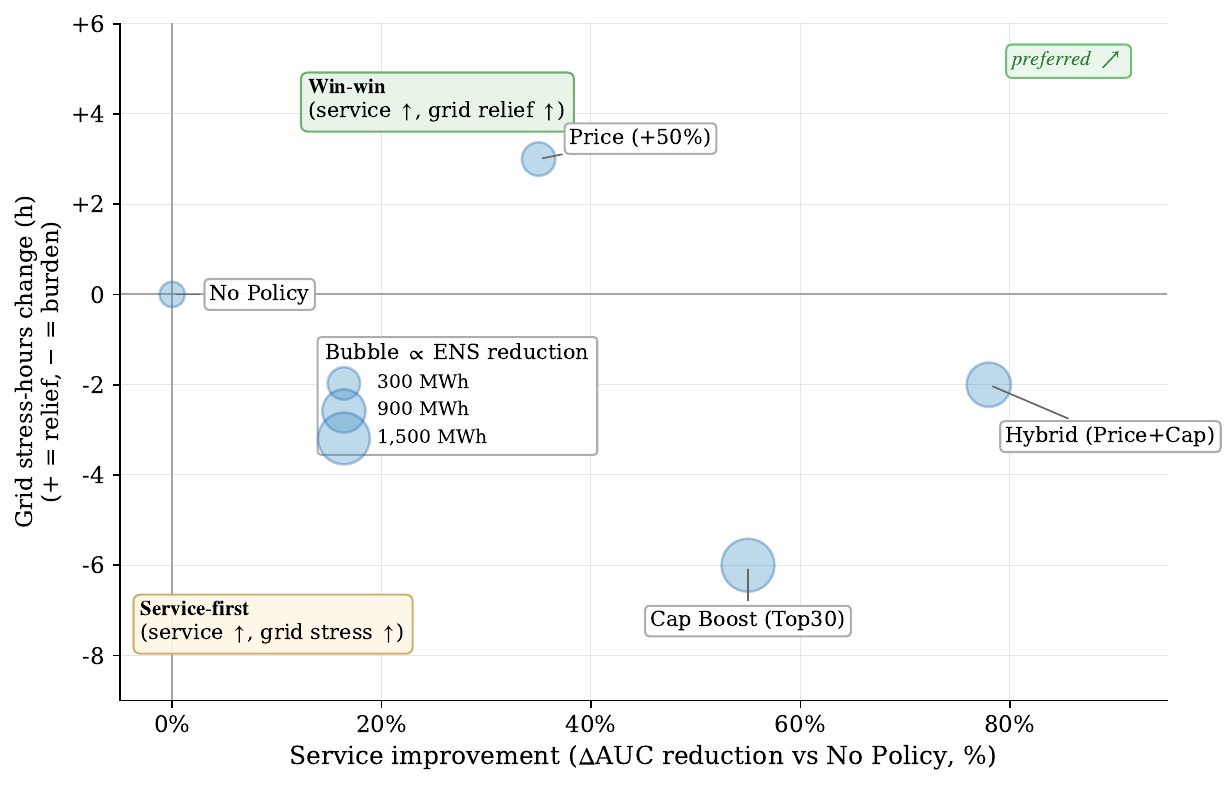}
\caption{Service--grid stress trade-off across policy interventions. Horizontal axis: service improvement ($\Delta$AUC reduction vs.\ No Policy); vertical axis: grid stress change (positive = relief, negative = burden). Price achieves win--win; CapBoost prioritizes service at grid cost; Hybrid attains near-Pareto performance. Bubble size indicates ENS reduction magnitude. The dashed line connecting policies approximates the Pareto frontier, with Hybrid closest to the ideal corner (high service improvement, low grid burden).}
\label{fig:fig10}
\end{figure}

Table~\ref{tab:policy} and Fig.~\ref{fig:fig10} summarize policy performance under representative \$m=1.5\$ stress. Price achieves 32.3\% backlog reduction with grid stress relief ($\Delta H_{stress}=+3$~h, indicating reduced stress). CapBoost achieves 55.4\% reduction but increases grid stress ($\Delta H_{stress}=-6$~h). Hybrid achieves 79.1\% reduction with super-additive synergy ($\sim1.8\times$ the sum of individual effects) while limiting grid penalty to 2~h.

The service--grid trade-off reveals distinct policy regimes: Price operates in the ``win--win'' zone (service improvement with grid relief); CapBoost prioritizes service at grid cost (``service-first'' zone); Hybrid achieves near-Pareto performance by combining demand smoothing with targeted capacity deployment.

\textbf{Interpretation of censored recovery times.} The $\dagger$ symbol in Table~\ref{tab:policy} indicates that recovery was not achieved within the 792-hour simulation window. For No Policy and Price, although $\Delta$AUC decreases (indicating reduced cumulative backlog), the recovery time remains censored because a persistent low-level ``tail'' of excess backlog remains above the recovery threshold $\theta$. This occurs when demand reduction slows backlog accumulation but does not accelerate clearance of existing backlogs---explaining why Price improves $\Delta$AUC but not $\Delta$RT. CapBoost and Hybrid achieve finite recovery times because capacity augmentation actively clears accumulated backlogs rather than merely slowing their growth.

\begin{table}[!t]
\centering
\caption{Policy comparison under representative demand shock (Exp1: $m=1.5$, 48-hour shock window).}
\label{tab:policy}
\footnotesize
\setlength{\tabcolsep}{4pt}
\begin{tabular}{@{}lccccc@{}}
\toprule
\textbf{Policy} & \textbf{$\Delta$AUC} & \textbf{$\Delta$RT} & \textbf{Peak} & \textbf{ENS} & \textbf{$\Delta H$} \\
 & \textbf{(M)} & \textbf{(h)} & \textbf{(k)} & \textbf{(GWh)} & \textbf{(h)} \\
\midrule
No Policy & 96.4 & 792$^\dagger$ & 145.4 & 206.39 & 0 \\
Price (+50\%) & 65.2 & 792$^\dagger$ & 95.5 & 205.93 & +3 \\
CapBoost & 43.0 & 661 & 134.8 & 204.91 & $-6$ \\
Hybrid & \textbf{20.2} & \textbf{451} & \textbf{85.9} & \textbf{204.46} & $-2$ \\
\bottomrule
\end{tabular}
\par\vspace{2pt}
{\scriptsize $^\dagger$Censored at 792\,h (unrecoverable within simulation window); see text for interpretation.}
\end{table}

\subsection{Resilience boundaries (RQ3)}

\begin{figure}[!t]
\centering
\includegraphics[width=\columnwidth]{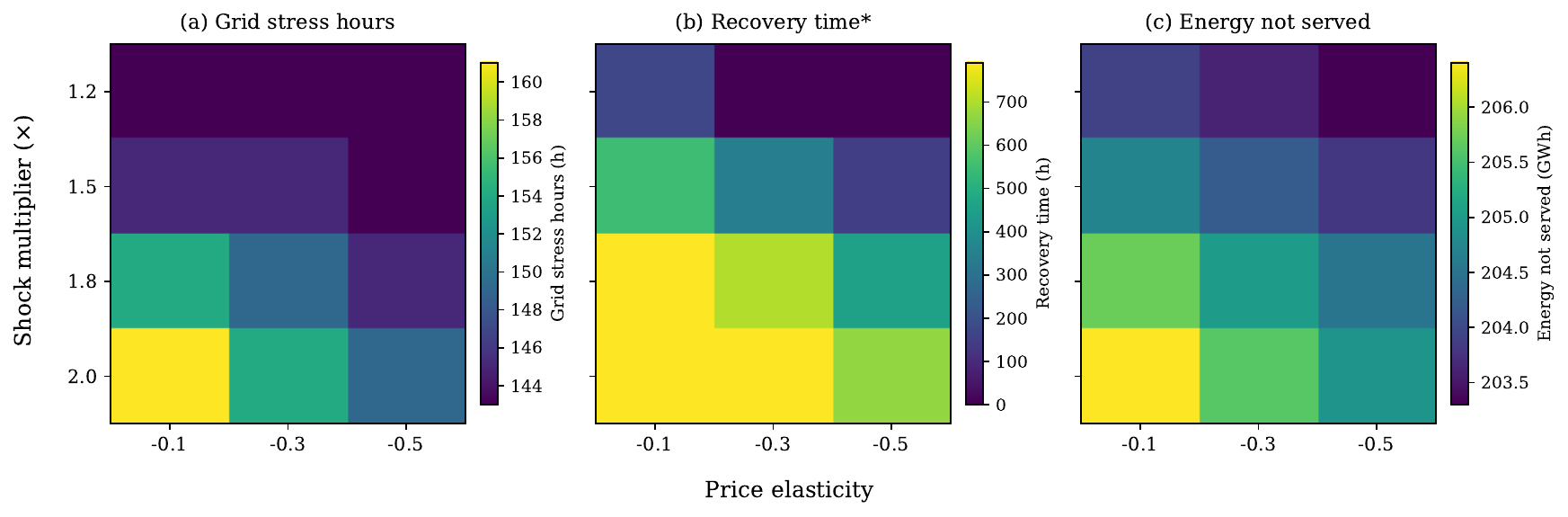}
\caption{Sensitivity analysis across stress multipliers and price elasticity. Panels show (a) grid stress hours, (b) recovery time, and (c) energy not served. Recovery time exhibits sharp phase transitions from recoverable to blocked states. Grid stress hours increase with demand stress and decrease with stronger demand response (more negative $\varepsilon$).}
\label{fig:fig11}
\end{figure}

Fig.~\ref{fig:fig11} presents sensitivity analysis across stress multipliers ($1.2\times$--$2.0\times$) and price elasticity ($\varepsilon=-0.1$ to $-0.5$). Recovery time exhibits sharp phase transitions between recoverable and blocked regimes. The blocking boundary varies systematically with elasticity: weakly elastic demand ($\varepsilon=-0.1$) becomes blocked at $m\ge1.8$; moderate elasticity ($\varepsilon=-0.3$) remains recoverable until $m=2.0$; strongly elastic demand ($\varepsilon=-0.5$) remains recoverable throughout the tested range.

This phase transition can be summarized by a conservative boundary approximation:
\begin{equation}
m_{crit}(\varepsilon) \approx 1.7 - 1.0\,\varepsilon,
\label{eq:resilience_boundary}
\end{equation}
where $m_{crit}$ represents the maximum absorbable stress multiplier for a given elasticity level. 

\textbf{Practical interpretation and caveats.} Equation~(\ref{eq:resilience_boundary}) provides an operator's rule-of-thumb: for a given achievable demand elasticity $\varepsilon$, the system can absorb stress shocks up to $m_{crit}$ while maintaining recoverability. For example, with $\varepsilon=-0.3$ (moderate price responsiveness), shocks up to $m\approx2.0$ remain recoverable; with $\varepsilon=-0.1$ (weak responsiveness), only shocks up to $m\approx1.8$ can be absorbed.

\textbf{Important limitations:} This boundary is empirically derived from our experimental range ($\varepsilon\in[-0.5,-0.1]$, $m\in[1.2,2.0]$) and should not be extrapolated beyond these bounds. The linear approximation may not hold for extreme elasticities or stress levels outside this range. Site-specific validation is recommended before operational deployment.

\begin{figure*}[!t]
\centering
\includegraphics[width=0.85\textwidth]{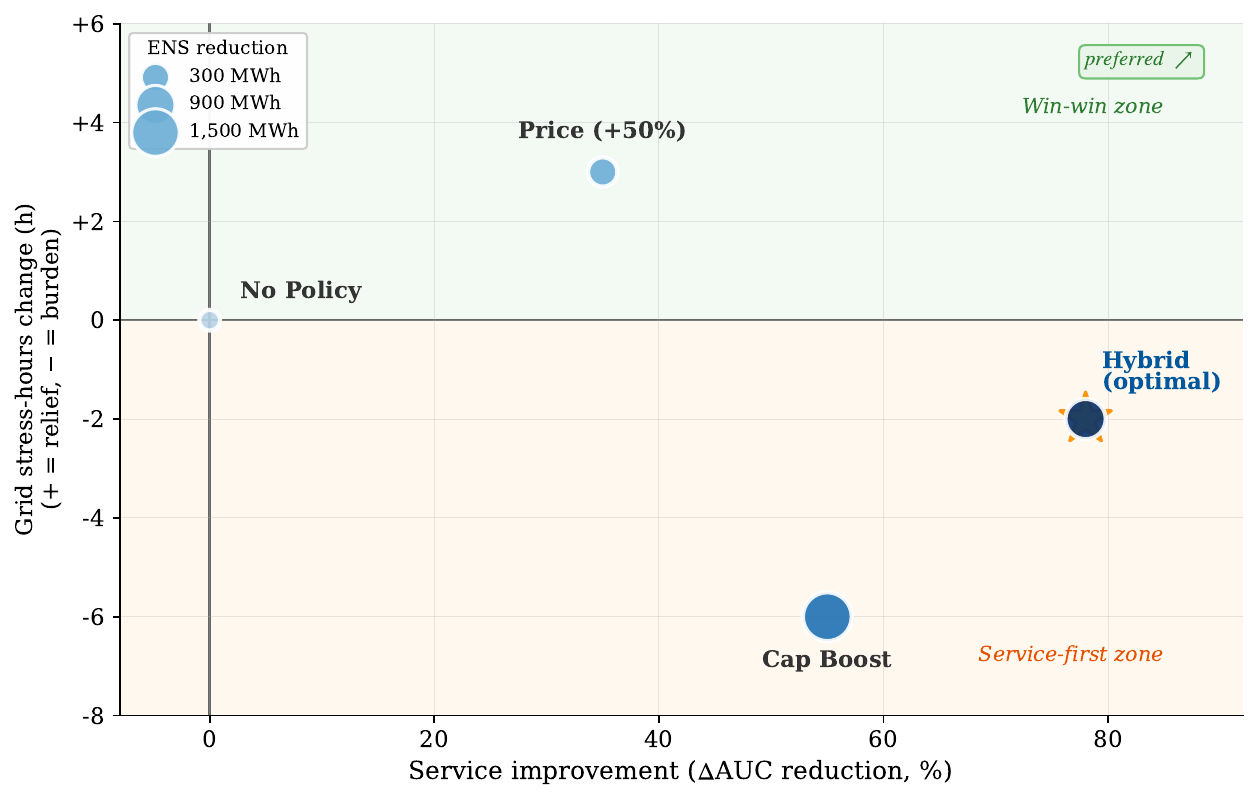}
\caption{Integrated service--grid trade-off summary. Each point represents a stress--policy scenario, plotting service improvement (backlog reduction) against grid impact (stress-hour change). The figure highlights near-Pareto regimes achievable through policy design and clarifies how intervention strategies shift the feasible resilience region.}
\label{fig:fig12}
\end{figure*}

Fig.~\ref{fig:fig12} provides an integrated service--grid trade-off summary consolidating scenario outcomes into a single frontier view for decision support.

\section{Discussion}\label{sec:discussion}

\subsection{Cross-scale physics injection as knowledge transfer}

The methodological core of this work is the cross-scale transfer function translating micro-level operational physics into city-scale risk characterization. Stage~1 learns a temperature- and pressure-conditioned deliverability surface from minute-resolution telemetry, encoding how station controllers allocate capacity under concurrent demand. Stage~2 injects this empirical law into the Shenzhen panel through anchored quantile mapping, ensuring that ``high stress'' in the macro domain corresponds to operating regimes actually observed at the micro level.

The ablation results (Table~\ref{tab:ablation} and Fig.~\ref{fig:fig9}) demonstrate why this transfer is essential rather than merely beneficial. A key finding is the ``lower error but wrong direction'' paradox: the non-injected baseline achieves nominally lower SLR prediction error yet exhibits counter-physical behavior---higher stress correlating with \emph{lower} predicted service loss---which would lead to systematically optimistic resilience assessments and potentially dangerous policy recommendations during extreme events. This finding aligns with broader Sci-ML literature suggesting that physics-informed constraints can prevent pathological extrapolation under distribution shift \cite{raissi2019physics,misyris2020physics,huang2023applications}. Our results provide empirical evidence that purely data-driven models trained on city-scale data may learn spurious correlations that fail precisely in the high-stress tail where accurate predictions matter most. Critically, this highlights that conventional forecasting metrics (RMSE, MAE) alone are insufficient for evaluating models intended for resilience planning; physics-consistency metrics must serve as a necessary complement.

\subsection{Service--grid trade-offs and policy implications}

The results reveal a fundamental trade-off between charging service resilience and distribution grid stress with important implications for coordinated infrastructure planning. Price-based demand shaping reduces arrivals during peak periods, lowering both instantaneous charging load and cumulative backlog \cite{wang2023shortterm,wang2025adaptive,kuang2024physics}. Capacity boosting takes the opposite approach, increasing service rate in targeted zones to accelerate backlog clearance, but concentrates power delivery in time, increasing transformer loading \cite{ray2023review,unterluggauer2022electric}.

The super-additive synergy in the Hybrid policy warrants mechanistic interpretation: price-based smoothing reduces peak demand during the shock window, lessening the peak load that capacity boosting must handle; simultaneously, capacity boosting accelerates clearance of residual backlogs that price intervention alone cannot eliminate. These findings suggest a practical two-layer operational strategy: deploy price signals broadly to smooth demand profiles while reserving targeted capacity interventions for localized hotspots.

\subsection{Implications for resilient distribution grid operation}

The derived resilience boundary $m_{crit}(\varepsilon)\approx 1.7-1.0\varepsilon$ has direct operational implications for distribution grid planning under extreme events. Grid operators can use this relationship to:
\begin{enumerate}
    \item Determine required demand flexibility (elasticity) to absorb anticipated stress multipliers during extreme weather or holiday surges.
    \item Identify zones where capacity augmentation is necessary when achievable elasticity falls below the resilience threshold.
    \item Design tiered emergency response protocols that activate price signals at lower stress levels and reserve capacity deployment for more severe scenarios.
\end{enumerate}

\subsection{Limitations and future directions}

Several limitations suggest directions for future research. First, the behavioral response model uses constant price elasticity; integrating user segmentation \cite{li2023electric} or agent-based models would strengthen realism. Second, Stage~5 grid coupling relies on transformer loading ratios rather than full AC power flow analysis; integration with detailed distribution network models would enable voltage and line-flow constraint evaluation. Third, cross-domain transfer is validated on a single source--target pair; broader validation across diverse urban contexts would strengthen generalizability claims. Finally, extending to full-year data would capture seasonal variations in temperature-dependent deliverability and stress patterns.

\section{Conclusion}\label{sec:conclusion}

This paper developed a five-stage cross-scale, physics-injected scientific machine learning framework for assessing urban EV charging network resilience under extreme demand stress while explicitly accounting for distribution grid constraints. The integrated pipeline comprises micro-scale deliverability learning from DC fast-charging telemetry (Stage~1), quantile-aligned physics injection into city-scale panels (Stage~2), dual-head spatio-temporal forecasting (Stage~3), backlog-based resilience simulation (Stage~4), and grid-coupled evaluation (Stage~5).

Three principal findings emerge. First, physics injection is essential for reliable extreme-event assessment, restoring the physically expected monotonic relationship between stress and service loss (Spearman $\rho=+1.0$ vs.\ $-0.8$ for non-injected baseline) and amplifying tail-regime sensitivity ($+923\%$ vs.\ $-6.8\%$ at $2.0\times$ stress). Notably, a ``lower error but wrong direction'' paradox emerges: the non-injected baseline achieves nominally lower SLR prediction error yet produces counter-physical stress response, highlighting that conventional forecasting metrics alone are insufficient for resilience assessment. Second, policy interventions induce quantifiable trade-offs: Price achieves 32.3\% backlog reduction with grid relief (win--win); CapBoost achieves 55.4\% reduction with increased grid stress (service-first); Hybrid achieves 79.1\% reduction with super-additive synergy. Third, resilience boundaries can be mapped via the empirical boundary $m_{crit}(\varepsilon)\approx 1.7-1.0\varepsilon$, linking demand flexibility to maximum absorbable stress.

The framework contributes methodologically by demonstrating how micro-scale physics can be systematically transferred to city-scale risk assessment; practically by providing quantified policy trade-offs bridging charging service planning and grid management; and operationally by offering resilience boundaries translating behavioral assumptions into actionable planning thresholds. These contributions directly address the need for advanced AI-driven, risk-aware operational planning methodologies for resilient distribution grids under extreme events.

\section*{Data Availability}

The Shenzhen UrbanEV dataset is publicly available \cite{li2025urbanev}. The Swiss charging station telemetry \cite{simolin2025analysis} is available open-source. Code for reproducing the Stage~1--Stage~5 pipeline will be made available upon publication at {\color{red}[GitHub repository link to be added]}.

\section*{CRediT Authorship Contribution Statement}

\textbf{Yifan Wang:} Conceptualization, Methodology, Software, Validation, Formal analysis, Investigation, Data curation, Writing -- original draft, Writing -- review \& editing, Visualization.

\section*{Declaration of Competing Interest}

The authors declare that they have no known competing financial interests or personal relationships that could have appeared to influence the work reported in this paper.

\section*{Declaration of Generative AI Use}

During the preparation of this work, the author used Claude (Anthropic) and ChatGPT (OpenAI) for assistance with manuscript formatting and language refinement. The author reviewed and edited the content as needed and takes full responsibility for the content of the published article.

\section*{Acknowledgements}

{\color{red}[To be completed. Please add funding information and any acknowledgements here.]}

\bibliographystyle{cas-model2-names-unsrt}
\bibliography{refs}

\end{document}